\documentclass{article}

\usepackage{PRIMEarxiv}

\usepackage{lineno,hyperref}
\usepackage{amsmath}
\usepackage{amsfonts}
\usepackage{amssymb}
\usepackage{geometry}
\usepackage{txfonts}
\usepackage{graphics}
\usepackage{graphicx}
\usepackage{bm}
\usepackage{booktabs}
\usepackage{multirow}
\usepackage{threeparttable}
\usepackage{caption}
\usepackage{subcaption}
\usepackage{arydshln}
\usepackage[utf8]{inputenc} 
\usepackage[T1]{fontenc}    
\usepackage{url}            
\usepackage{nicefrac}       
\usepackage{microtype}      
\usepackage{fancyhdr}       

\newcommand{\brc}[1]{\left(#1\right)}

\usepackage[numbers]{natbib}
\usepackage{setspace}

\title{Physics Informed Neural Networks for Free Shear Flows}

\author{Siddharth Raghu\\
  Department of Applied Mechanics\\
  Indian Institute of Technology Delhi\\
  \And
      Rajdip Nayek \\
  Department of Applied Mechanics\\
  Indian Institute of Technology Delhi\\
  \texttt{rajdipn@am.iitd.ac.in} \\
  \And
      Vamsi K Chalamalla \\
  Department of Applied Mechanics\\
  Indian Institute of Technology Delhi\\
  \texttt{vchalama@am.iitd.ac.in} \\
}

\begin{document}
\maketitle

\begin{abstract}
The transformative impact of machine learning, particularly Deep Learning (DL), on scientific and engineering domains is evident. In the context of computational fluid dynamics (CFD), Physics-Informed Neural Networks (PINNs) represent a significant innovation, enabling data-driven fluid simulations while incorporating physics-based laws described by partial differential equations (PDEs). While PINNs have demonstrated efficacy in various fluid flow scenarios, a noticeable gap exists in their application to simulate jet flows—an essential category in engineering. Jets, crucial for downburst outflow, ventilation, and heat transfer, lack comprehensive exploration through PINNs in existing literature.

This study addresses this gap by focusing on the application of PINNs to simulate steady jet flows, specifically 2D planar turbulent jet flow scenarios. The novelty lies not only in adapting PINNs for simulating jet flows but also in overcoming challenges such as poor convergence during training, attributed to imbalances in loss terms. We propose a novel PINN architecture for Reynolds-Averaged Navier–Stokes (RANS) simulation of steady turbulent jet flows, without the need for turbulence models and simulation data. Additionally, an extended dynamic weighting strategy is introduced to enhance the balance of loss term contributions in the PINN, resulting in improved convergence and more accurate predictions. 
\end{abstract}


\section{Introduction}
The recent surge in machine learning (ML) has significantly influenced various scientific and engineering fields \cite{RAISSI2019686}. Deep Learning (DL), as a pivotal component of this ML wave, has found applications across diverse fields, including astronomy \cite{waldmann2019mapping}, climate modeling \cite{ham2019deep}, solid mechanics \cite{mianroodi2021teaching}, and chemistry \cite{segler2018planning}. In the context of scientific applications related to computational fluid dynamics (CFD), two predominant approaches existed:  \textit{physics-driven} methods relying on traditional numerical solvers for solving physical laws expressed through coupled partial differential equations \cite{russell1983finite, brooks1982streamline, karniadakis2005spectral}, and \textit{data-driven} methods, particularly deep learning, capable of constructing cheap surrogate models of flow phenomena when trained with ample data.

However, generating training data for deep neural networks typically involves time-intensive simulations of fluid dynamics using numerical solvers. 
With the rise of modern deep neural networks, the ML community became interested in the potential of conducting data-driven fluid simulations without extensive training data, while still incorporating physics-based laws described by PDEs. Physics-informed neural networks (PINNs) have emerged as a class of physics-informed-deep-learning approaches that address these requirements. At their core, PINNs feature a deep neural network that approximates PDE solutions, effectively transforming the problem of solution approximation into one focused on minimizing a loss function. This composite loss function includes terms addressing initial and boundary conditions, as well as the PDE residual at specific collocation points within the domain. Post-training, the network can generate solutions for a given input point within the integration domain of a differential equation. What sets PINNs apart from generic neural networks is the incorporation of a physics-based residual error in the loss function. This inclusion represents a noteworthy innovation, enhancing the network's ability to capture and integrate the underlying physics, eliminating the need for training data derived from previous simulations or experiments. 

The exploration of PINNs has given rise to several innovative deep architectures, offering promising applications in the simulation of fluid dynamics.
Noteworthy studies by Raissi et al.\ \cite{RAISSI2019686} demonstrated the efficacy of PINNs in addressing laminar and turbulent flows across various scenarios, including cylinder flow, 2D channel flow with a spherical obstacle, and 3D intracranial aneurysm. Sun et al.\ \cite{sun2020surrogate} contributed to PINN development by introducing hard boundary conditions and experimenting with activation functions and adaptive learning rates, resulting in PINNs demonstrating excellent agreement with CFD simulations, particularly in predicting aneurysmal flow. 
The integration of convolutional neural networks into PINNs, as demonstrated by Zhu et al.\ \cite{zhu2019physics}, showcased accurate predictions for 2D Darcy flow, surpassing the performance of conventional fully-connected PINNs. Jin et al.\ \cite{jin2021nsfnets} highlighted the versatility of PINNs in handling diverse flow scenarios, spanning Kovasznay, cylinder wake, Beltrami, and turbulent channel flows. In the realm of additive manufacturing, Zhu et al.\ \cite{zhu2021machine} applied PINNs with hard boundary-constraint enforcement to predict temperature and dynamics in melt pool fluid dynamics, illustrating their capability to address complex, dynamic systems with moderate datasets.  
Exploiting the potential of CNN-based PINNs, Wang et al.\ \cite{wang2020towards} introduced TF-Net for accurate 2D multiscale turbulent flow predictions, emphasizing PINNs' ability to capture multiscale phenomena in fluid dynamics. Pioch et al.\ \cite{pioch2023turbulence} effectively employed PINNs to reconstruct recirculating flow over a backward step, a well-known problem in fluid dynamics. Lucor et al.\ \cite{LUCOR2022111022} used PINNs for the forward solution of 3D Rayleigh-B\'{e}nard convection within a cavity, achieving reasonably accurate predictions for the flow. However, simulating turbulent flows using instantaneous Navier-Stokes equations poses computational challenges. Addressing this, Eivazi et al. \cite{eivazi2022physics} proposed an alternative approach by employing Reynolds-Averaged Navier-Stokes (RANS) equations without turbulence models in PINNs to predict turbulent flows. Their methodology involved predicting not only velocities and pressure but also Reynolds stresses, proving effective across various turbulent flow fields, from boundary layer flows to airflow over an airfoil, achieving excellent predictions. Accurate forward solutions with PINNs have been noted to be more difficult to obtain compared to inverse PINNs where data also acts as a regularizer.

Despite these advancements, there is a critical gap in exploring PINNs for simulating jet flows --- a significant category of free-shear flows with paramount engineering significance. Jets are pivotal in scenarios such as downburst outflow impact on structural components, ventilation systems, and heat transfer processes. Many studies have investigated turbulent jets through experimental and numerical approaches \cite{rhea2009rans,jaramillo2012dns,yan2018evaluation,khayrullina2019validation,agrawal2023numerical,deo2008influence,stanley2002study}. Despite their importance, the literature lacks a comprehensive utilization of PINNs for simulating jet flows.

In many engineering applications, especially those involving steady flows, the Reynolds-averaged Navier–Stokes (RANS) equations offer a suitable framework for fluid simulation. Consequently, our study is geared towards addressing this critical gap by honing in on the potential of PINNs in simulating steady jet flows. Specifically, our focus lies on 2D planar jet flow scenarios, chosen for their simplified yet representative nature. This strategic choice enables an effective exploration of PINNs' capabilities in capturing the underlying physics of jet flows.

It is well known that vanilla PINNs, as introduced by Raissi et al.\ \cite{RAISSI2019686}, face limitations detailed in \cite{cai2021physics}. A significant challenge arises from the poor convergence of PINNs during training, stemming from \textit{imbalances in the loss terms}. In fluid dynamics, governed by the nonlinear Navier-Stokes (NS) equations --- which is a system of multiple coupled PDEs based on mass and momentum conservation principles ---, the loss function for training PINNs includes multiple residual terms, each corresponding to one PDE, along with contributions from boundary and initial conditions. Imbalances in loss terms occur when the magnitude of one loss term significantly outweighs others, leading to premature convergence in the composite loss function. Although these individual losses are typically combined through a weighted summation with static weights -- kept fixed throughout training --, tuning these weights proves challenging. Wang et al.\ \cite{wang2021understanding} observed substantial differences in the gradient values of each constraint during backpropagation, which contributes to the imbalance. To tackle this issue, they proposed a method of dynamically adjusting the weights between different components in the loss to improve training convergence. 

In the pursuit of simulating steady jet flows using PINNs, this paper introduces a novel PINN architecture for forward RANS simulation of a steady turbulent jet flow without reliance on a turbulence model and simulation data. Additionally, an extended dynamic weighting strategy is proposed to better balance the contribution of each loss term in the PINN. Inspired by the dynamic weighting scheme of \cite{wang2021understanding}, this strategy uses trainable weights for each PDE, updated based on the gradient information of the corresponding loss term. The proposed architecture, coupled with the extended dynamic weighting strategy, enhances loss convergence and yields more accurate predictions compared to PINN models previously proposed by Eivazi et al.\ \cite{eivazi2022physics} and Pioch et al.\ \cite{pioch2023turbulence}.

\section{Free Shear Flow: Planar Jet Flow}
\subsection{Simulation Setup}

\begin{figure}[htpb]
    \centering
    \includegraphics[clip, trim=0cm 0cm 0cm 0cm,scale=0.6]{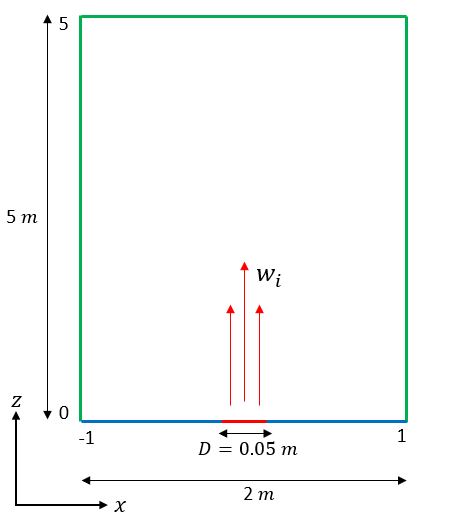}
    \caption{A schematic of the computational domain for the present study. The top and side boundaries are considered to be
    outlet boundaries. The bottom boundary is considered to be a wall except at the center where the jet fluid enters the computational domain.}
    \label{fig:domain}
\end{figure}

In this study, we conduct two-dimensional simulations of steady laminar and turbulent planar jets using different architectures of PINN. The spatial configuration of the problem is shown in Fig~\ref{fig:domain}, where the domain extends from -1 m to 1 m in the radial direction (x-axis) and from 0 to 5 m in the axial direction (z-axis). A continuous stream of the jet is introduced at the bottom center into a quiescent environment of fluid. The jet width of 0.05 meters is maintained constant while the inlet velocity of the jet is varied to obtain a range of Reynolds numbers spanning from laminar to fully turbulent regime. The simulation domain for the jet was chosen to align with the domain used by Agarawal et.\ al.\ \cite{agrawal2023numerical}. Due to the utilization of a uniform mesh in this study, the jet width was increased to ensure an adequate distribution of collocation points along the jet entrance. Consequently, the axial extent of the simulation domain was also enlarged to capture the self-similar region of the jet. The domain is discretized into a uniform mesh with 1000 points in the axial direction and 400 points in the radial direction, and this mesh remains consistent across all Reynolds numbers considered for the jet simulations. The selection of the mesh was based on the initial examination of the mesh dependence in PINN results for the jet. Due to the absence of a precise correlation between mesh refinement and PINN accuracy, the decision was made to adopt a mesh similar to that used in CFD simulations. The working fluid chosen for these simulations is water, characterized by a density ($\rho$) of $1000$ $kg/m^3$ and a dynamic viscosity ($\mu$) of $0.001 Pa.s$.

\subsection{Theoretical Background}

The planar laminar and turbulent jet flow considered in this study are governed by two-dimensional incompressible Navier-Stokes equations and Reynolds-Averaged Navier-Stokes equations respectively. In PINN simulations for the laminar jet, the flow variables are scaled to non-dimensional values as follows:
\begin{equation}
    \label{eq:ND_PINN}
    u^* = \frac{u}{w_{i}}, \; w^* = \frac{w}{w_{i}}, \; p^* = \frac{p}{\rho w_{i}^2}, \; x^* = \frac{x}{D}, \; z^* = \frac{z}{D}
\end{equation}
Here, $w_{i}$ is inlet velocity and $D$ is inlet width of the jet. Here, $u^*,w^*,p^*$ are non-dimensional quantities, and $u,w,p$ are dimensional quantities. For simplicity, the asterisks are dropped for non-dimensional quantities for the rest of the manuscript. Following are the non-dimensional Navier-Stokes equations in two-dimensional space:
\begin{align}
\label{eq:NS_ND_C}
    \frac{\partial u}{\partial x} + \frac{\partial w}{\partial z} \; &= \; 0, \\
\label{eq:NS_ND_X}
    u\frac{\partial u}{\partial x} + w\frac{\partial u}{\partial z} &= -\frac{\partial p}{\partial x} + \frac{1}{\textit{Re}}\brc{\frac{\partial^2 u}{\partial x^2} + \frac{\partial^2 u}{\partial z^2}}, \\
\label{eq:NS_ND_Z}
    u\frac{\partial w}{\partial x} + w\frac{\partial w}{\partial z} &= -\frac{\partial p}{\partial z} + \frac{1}{\textit{Re}}\brc{\frac{\partial^2 w}{\partial x^2} + \frac{\partial^2 w}{\partial z^2}}.
\end{align}
The Reynolds number of the jet is defined using the inlet velocity and the jet width given by $Re=\rho w_i D/\mu$. For the turbulent jet simulations, we consider Reynolds Averaged Navier Stokes equations (RANS) as we are primarily interested in predicting the mean flow statistics of the jet using PINN. RANS is a widely used tool for studying turbulent flows at a much smaller computational cost when compared to more elaborate large-eddy simulations and direct numerical simulations. However, modeling the unclosed Reynolds stress terms in RANS equations remains a significant challenge to date. In this study, we utilize PINN to find the solution of turbulent jet flow, using RANS equations, without any underlying turbulence models to model the Reynolds Stress terms. RANS variables are non-dimensionalized as shown below:
\begin{equation}
    \label{eq:ND_RS_PINN}
    \overline{u^*} = \frac{\overline{u}}{{w}_{i}}, \; \overline{w^*} = \frac{\overline{w}}{{w}_{i}}, \; \overline{p^*} = \frac{\overline{p}}{\rho{{w}_{i}}^2}, \; x^* = \frac{x}{D}, \; z^* = \frac{z}{D}, \; \overline{{u'u'}^*} = \frac{\overline{u'u'}}{{w_{i}}^2}, \; \overline{{u'w'}^*} = \frac{\overline{u'w'}}{{w_{i}}^2}, \; \overline{{w'w'}^*} = \frac{\overline{w'w'}}{{w_{i}}^2}.
\end{equation}
Here, the overbar for each variable represents the time-averaged variable. The corresponding non-dimensional RANS equations after dropping the asterisk are shown below:
\begin{align}
\label{eq:RANS_ND_C}
    \frac{\partial \overline{u}}{\partial x} + \frac{\partial \overline{w}}{\partial z} \; &= \; 0, \\
\label{eq:RANS_ND_X}
    \overline{u}\frac{\partial \overline{u}}{\partial x} + \overline{w}\frac{\partial \overline{u}}{\partial z} &= -\frac{\partial \overline{p}}{\partial x} + \frac{1}{\textit{Re}}\brc{\frac{\partial^2 \overline{u}}{\partial x^2} + \frac{\partial^2 \overline{u}}{\partial z^2}} - \frac{\partial \overline{u'u'}}{\partial x} - \frac{\partial \overline{u'w'}}{\partial z}, \\
\label{eq:RANS_ND_Z}
    \overline{u}\frac{\partial \overline{w}}{\partial x} + \overline{w}\frac{\partial \overline{w}}{\partial z} &= -\frac{\partial \overline{p}}{\partial z} + \frac{1}{\textit{Re}}\brc{\frac{\partial^2 \overline{w}}{\partial x^2} + \frac{\partial^2 \overline{w}}{\partial z^2}} - \frac{\partial \overline{u'w'}}{\partial x} - \frac{\partial \overline{w'w'}}{\partial z}.
\end{align}
The Reynolds stress terms are usually modelled using a turbulence model with the help of Boussinesq hypothesis as shown in Equation~\ref{eq:bh_aprx}
\begin{equation}
    \label{eq:bh_aprx}
    \overline{u_{i}'u_{j}'} = - \nu_{t}\brc{\frac{\partial \overline{u_{i}}}{\partial x_{j}} + \frac{\partial \overline{u_{j}}}{\partial x_{i}}} - \frac{2}{3}k\delta_{ij}
\end{equation}
Where $k$ is the turbulent kinetic energy, which can be described as below for the 2D incompressible flow, and $\delta_{ij}$ is the Kronecker Delta.
\begin{equation}
    \label{eq:k}
    k = \frac{1}{2}\overline{u_{i}'u_{i}'}
\end{equation}
Substituting the expression for $k$ in the Equation \ref{eq:bh_aprx} and solving for the Reynolds Stress terms, the Boussinesq hypothesis reduces to Equation \ref{eq:bh}.
\begin{equation}
    \label{eq:bh}
    \overline{u_{i}'u_{j}'} = - \nu_{t}\brc{\frac{\partial \overline{u_{i}}}{\partial x_{j}} + \frac{\partial \overline{u_{j}}}{\partial x_{i}}}
\end{equation}
Extensive research on planar jets has been conducted in the past, as evident by studies such as those by Gutmark et.\ al.\, Heskestad et.\ al.\, and Bradbury et.\ al.\ \cite{gutmark1976planar, heskestad1965hot, bradbury1965structure}. These investigations aimed to ascertain flow statistics, and collectively, they established that self-similarity in a planar jet is attained after an axial distance of $30 \times D$. Thus, for a laminar jet, it is now well established that the centerline velocity and the half-width of the jet as a function of $z$, as described in Equation~\ref{eq:lam_w0} and \ref{eq:lam_l1/2}, in the self-similar region. 
\begin{align}
\label{eq:lam_w0}
    w_{0}(z) &\propto z^{-\frac{1}{3}}\\
\label{eq:lam_l1/2}
    l_{\frac{1}{2}}(z) &\propto z^\frac{2}{3}
\end{align}
Where, $w_{0}$ is the center-line velocity of the laminar jet, $l_{\frac{1}{2}}$ is the half-width of the jet computed from axial velocity of the laminar jet. The corresponding self-similar profiles for a turbulent jet are given in the Equation~\ref{eq:turb_w0} and \ref{eq:turb_l1/2}.
\begin{align}
\label{eq:turb_w0}
    \overline{w}_{0}(z) &\propto z^{-\frac{1}{2}}\\
\label{eq:turb_l1/2}
    \overline{l}_{\frac{1}{2}}(z) &\propto z
\end{align}
Where, $\overline{w}_{0}$ is the mean center-line velocity of the turbulent jet, $\overline{l}_{\frac{1}{2}}$ is the half-width of the jet computed from mean axial velocity of the turbulent jet.

The above mentioned flow statistics will be used in this study to evaluate the accuracy of flow predictions by the PINNs for both laminar and turbulent jets.

\subsection{CFD Simulations}

To compare the results obtained from PINN, CFD simulations using two-dimensional RANS equations are performed for different values of Reynolds numbers. To validate the accuracy of benchmark CFD simulations, the results were validated against the established theory of jets \cite{pope2000turbulent,kundu2015fluid}. 

The CFD simulations were set up using Ansys FLUENT in a 2D computational domain, with domain and boundary conditions consistent with PINN configuration. For the laminar jet, a laminar flow solver was adopted, while for the turbulent jet, the standard $k-\epsilon$ turbulence model was employed to model Reynolds stresses. The parameters of the $k-\epsilon$ model were selected based on previous studies \cite{rhea2009rans,jaramillo2012dns,yan2018evaluation,khayrullina2019validation,agrawal2023numerical}. The simulations were performed until the residuals of all the equations reached below $10^{-6}$.The boundary conditions for laminar and turbulent jets are outlined in Tables~\ref{tab:boundary_conditions} and \ref{tab:turbulent_boundary_conditions} respectively.

\begin{table}[h!]
  \caption{Boundary conditions for laminar jet simulations.}
  \begin{center}
  \begin{tabular}{ | c | c | c | c | }
    \hline
    \textbf{Variables}  &  \textbf{Inlet} & \textbf{Outlet} & \textbf{Wall} \\ 
    \hline 
    u &  Zero & Zero-gradient & Zero \\
    w &  $w_i$ & Zero-gradient & Zero \\
    p &  Zero & Zero-gradient & Zero-gradient \\
    \hline
  \end{tabular}
  \label{tab:boundary_conditions}
  \end{center}
\end{table}

\begin{table}[!h]
  \caption{Boundary conditions for turbulent jet simulations.}
  \begin{center}
  \begin{tabular}{ | c | c | c | c | }
    \hline
    \textbf{Variables}  &  \textbf{Inlet} & \textbf{Outlet} & \textbf{Wall} \\ 
    \hline 
    $\overline{u}$ &  Zero & Zero-gradient & Zero \\
    $\overline{w}$ &  $w_i$ & Zero-gradient & Zero \\
    $\overline{p}$ &  Zero & Zero-gradient & Zero-gradient \\
    $\overline{u'u'}$ &  Zero & Zero-gradient & Zero \\
    $\overline{u'w'}$ &  Zero & Zero-gradient & Zero \\
    $\overline{w'w'}$ &  Zero & Zero-gradient & Zero \\
    \hline
  \end{tabular}
  \label{tab:turbulent_boundary_conditions}
  \end{center}
\end{table}

\section{Methodology: PINN}
\subsection{PINN Architectures}

\begin{figure}[htpb]
    \centering
    \includegraphics[width=1\linewidth]{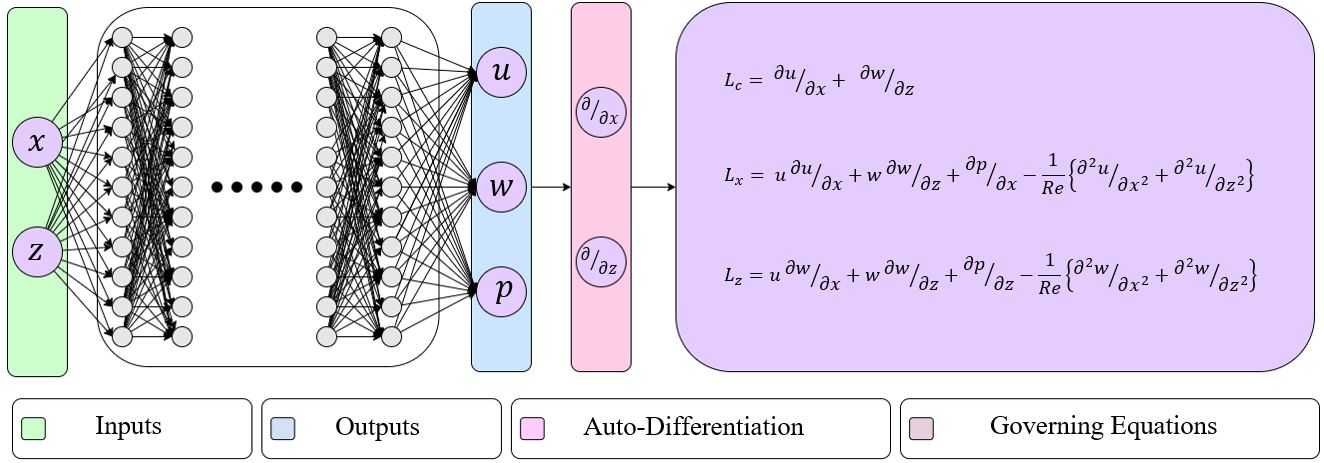}
    \caption{A schematic of PINN for solving Navier-Stokes equations for steady laminar jet.}
    \label{fig:pinn_laminar}
\end{figure}

For PINN solutions to the laminar jet case, the neural network architecture comprises $100 \times 10$ hidden layers, taking two spatial inputs ($x, z$) and producing three outputs ($u, w, p$), as illustrated in Fig~\ref{fig:pinn_laminar}.

For turbulent jets, four different architectures have been analysed:

\begin{itemize}

\item[\textbf{A1}]: In this case, $l_{m}$ is an additional output of the neural network, alongside the three outputs ($u, w, p$). The architecture, shown in Fig~\ref{fig:pinn_turbulent_ml}, includes two spatial inputs ($x, z$) and four flow variables ($\overline{u}, \overline{w}, \overline{p}, l_{m}$) as outputs, with $100 \times 10$ hidden layers. The governing equations include only the dominant Reynolds stress gradient term ($\overline{u'w'}$), as described in the theory of jets, modeled using the Prandtl's mixing length model (Equations ~\ref{eq:Ml_model} and \ref{eq:t_vis}).
\begin{align}
\label{eq:Ml_model}
    \overline{u'w'} = -\nu_{t}\frac{\partial \overline{w}}{\partial x}\\
\label{eq:t_vis}
    \nu_{t} = l_{m}^2 \left|\frac{\partial \overline{w}}{\partial x} \right|
\end{align}

\begin{figure}[htpb]
    \centering
    \includegraphics[width=1\linewidth]{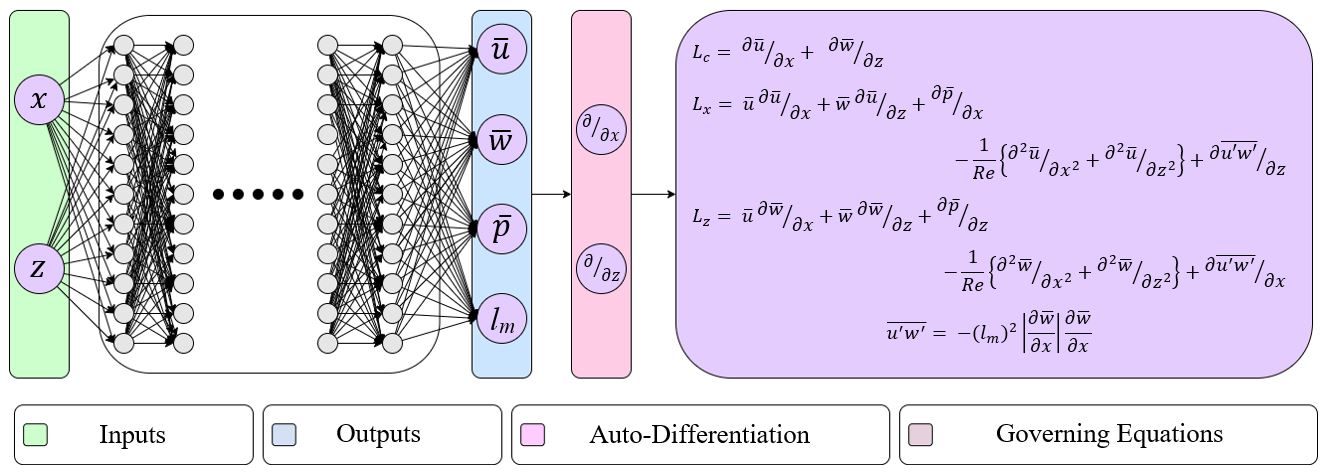}
    \caption{A schematic of PINN for solving RANS equations, with only $\overline{u'w'}$ Reynolds stress term which is modeled using Prandtl's mixing length model, with mixing length, $l_m$, as an output of the neural network, for the steady turbulent jet.}
    \label{fig:pinn_turbulent_ml}
\end{figure}

\item[\textbf{A2}]: In this case, the PINN architecture is presented in Fig~\ref{fig:pinn_RANS}, with two spatial inputs ($x, z$) and six flow variables ($\overline{u}, \overline{w}, \overline{p}, \overline{u'u'}, \overline{u'w'}, \overline{w'w'}$) as outputs and $100 \times 10$ hidden layers. The governing equations are based on RANS equations (Equations~\ref{eq:RANS_ND_C}, \ref{eq:RANS_ND_X}, and \ref{eq:RANS_ND_Z}), with Reynolds stress terms as additional outputs of the neural network. This architecture was introduced by Eivazi et.\ al.\ \cite{eivazi2022physics}.

\begin{figure}[htpb]
    \centering
    \includegraphics[width=1\linewidth]{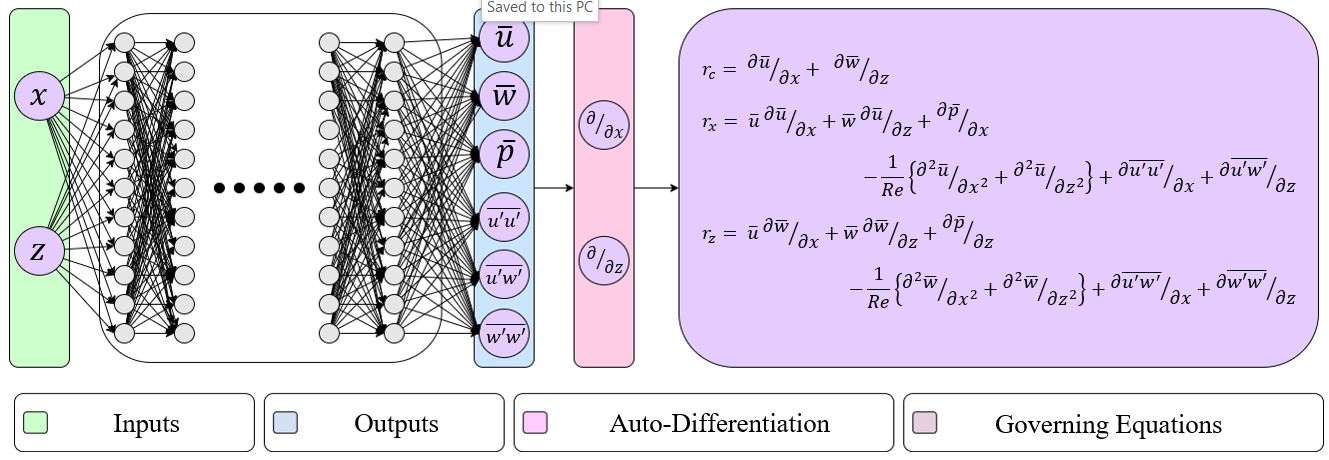}
    \caption{A schematic of PINN for solving RANS equations, with the Reynolds stresses being the output of the neural network, for the steady turbulent jet.}
    \label{fig:pinn_RANS}
\end{figure}

\item[\textbf{A3}]: In this case, the Boussinesq hypothesis is utilized which relates the Reynolds stresses to mean velocity gradients with the turbulent viscosity as the proportionality constant. The PINN architecture, shown in Fig~\ref{fig:pinn_nut}, includes two spatial inputs ($x, z$) and four flow variables ($\overline{u}, \overline{w}, \overline{p}, \nu_t$) as outputs, with $100 \times 10$ hidden layers. This architecture was utilized by Pioch et.\ al.\ \cite{pioch2023turbulence}. The governing equations are based on RANS equations (Equations~\ref{eq:RANS_ND_C}, \ref{eq:RANS_ND_X}, \ref{eq:RANS_ND_Z}) and the Boussinesq hypothesis (Equation~\ref{eq:bh}).

\begin{figure}[htpb]
    \centering
    \includegraphics[width=1\linewidth]{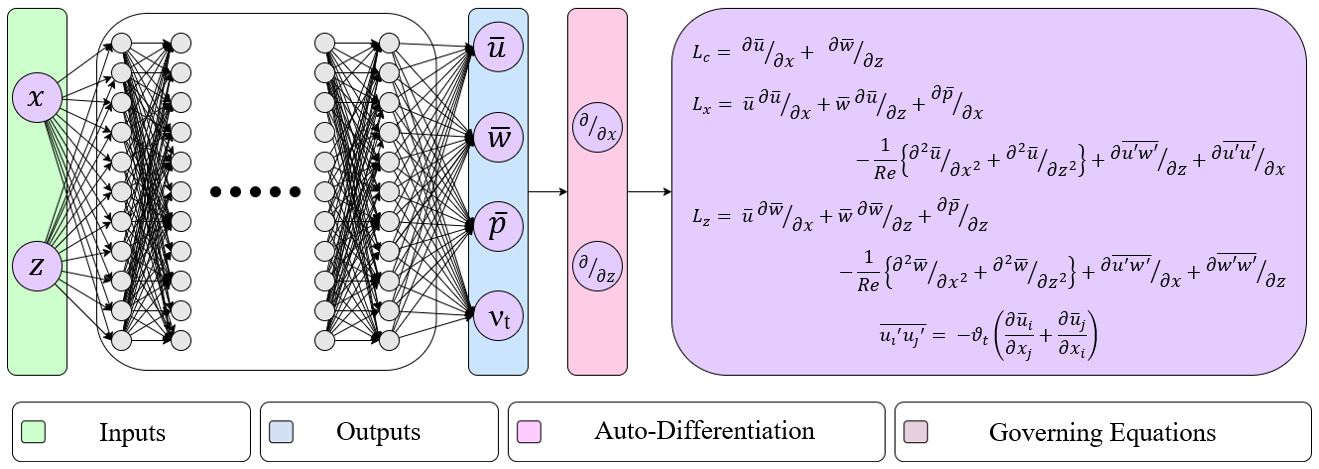}
    \caption{A schematic of PINN for solving RANS equations, with Reynolds stresses being modeled using the Boussinesq hypothesis where the turbulent viscosity is also considered as part of the PINN outputs, for the steady turbulent jet.}
    \label{fig:pinn_nut}
\end{figure}

\item[\textbf{A4}]: In this case, the Boussinesq hypothesis is utilized as in A3. The PINN architecture, shown in Fig~\ref{fig:pinn_2NN_nut}, includes two spatial inputs ($x, z$) and three flow variables ($\overline{u}, \overline{w}, \overline{p}$) as outputs, with $100 \times 10$ hidden layers. In contrast to A3, it incorporates a second neural network with four velocity gradients as inputs and turbulent viscosity ($\nu_{t}$) as output, with $50 \times 5$ hidden layers. The governing equations are based on RANS equations (Equations~\ref{eq:RANS_ND_C}, \ref{eq:RANS_ND_X}, \ref{eq:RANS_ND_Z}) and the Boussinesq hypothesis (Equation~\ref{eq:bh}).

\begin{figure}[htpb]
    \centering
    \includegraphics[width=1\linewidth]{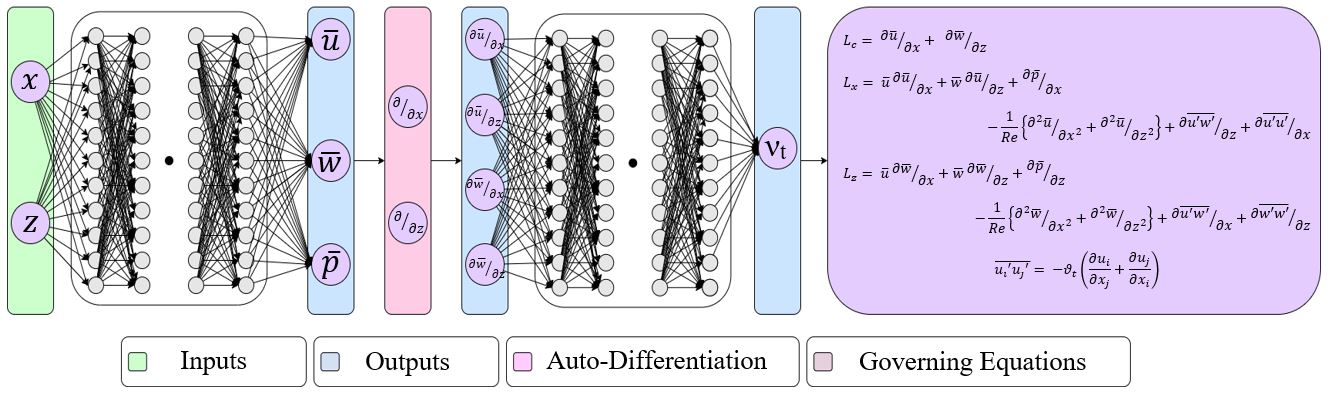}
    \caption{A schematic of PINN for solving RANS equations, with Reynolds stresses being modeled using the Boussinesq hypothesis where the turbulent viscosity is made to be a function of velocity gradients using a second neural network, for the steady turbulent jet.}
    \label{fig:pinn_2NN_nut}
\end{figure}
\end{itemize}

\subsection{Composite Loss Function and Training of PINN}
The loss function in this study is the Mean Squared Error (MSE) loss for training the neural networks. The total loss function ($L$) is a combination of the PDE loss ($L_{PDE}$) and boundary loss ($L_b$) with their respective weights ($\alpha$), as shown in Equation~\ref{eq:loss}. The gradients in the PDE losses are computed using the automatic differentiation functionality of PyTorch.

\begin{align}
\label{eq:loss}
    L &= L_{PDE} + \alpha L_{b}\\
\label{eq:PDE_loss}
    L_{PDE} &= L_{c} + L_{x} + L_{z}
\end{align}
where, $L_{PDE}$ is the residual loss of Navier-Stokes/RANS equations, as defined in  Equation~\ref{eq:PDE_loss}.The total PDE loss is the sum of losses from continuity ($L_c$), x-momentum ($L_x$), and z-momentum ($L_z$) equations defined by Equations~\ref{eq:NS_ND_C}, \ref{eq:NS_ND_X} and \ref{eq:NS_ND_Z} respectively. $L_{PDE}$ is a combination of residual loss of different equations in Navier-Stokes/RANS equation as shown in 

Typically, a gradient descent algorithm is employed to train neural network parameters using the following update rule:
\begin{equation}
    \label{eq:GSD}
    {\theta}^{k+1} = {\theta}^{k} - \eta{\nabla}_{\theta}L_{PDE} - \eta\alpha{\nabla}_{\theta}L_{b}.
\end{equation}
Here, $k$ denotes the current iteration step, ${\theta}$ represents the neural network parameters (weights and biases), $\eta$ is the learning rate, ${\nabla}_{\theta}L_{PDE}$ is the gradient of PDE loss with respect to neural network parameters, and ${\nabla}_{\theta}L_{b}$ is the gradient of boundary loss with respect to the neural network parameters.

The parameter $\alpha$ represents the boundary loss weight and is typically kept constant (or static) in standard PINN configuration. It serves as a hyper-parameter that requires careful tuning for good results. However, maintaining a static weight through the training iterations poses a challenge of gradient pathologies, leading to inaccurate results in the case of forward PINN. To overcome this issue, Wang et al. \cite{wang2021understanding, JIN2021109951} proposed a dynamic weight strategy, expressed as follows:
\begin{align}
\label{eq:dw_alpha}
    \hat{\alpha}^{k+1} &= \frac{\overline{|{\nabla}_{\theta} L_{PDE}|}}{\overline{|{\nabla}_{\theta} L_{b}|}}\\
\label{eq:dw_lambda}
    \alpha^{k+1} &= (1-\lambda)\alpha^{k} + \lambda\hat{\alpha}^{k+1}
\end{align}
Here, $k$ is the current iteration step, ${\theta}$ represents the neural network weights, $\lambda$ is a hyperparameter defining the contribution from the previous weights, $\overline{|{\nabla}_{\theta} L_{PDE}|}$ is the mean of $|{\nabla}_{\theta} L_{PDE}|$, and $\overline{|{\nabla}_{\theta} L_{b}|}$ is the mean of $|{\nabla}_{\theta} L_{b}|$.

Given that the PDE loss $L_{PDE}$ inherently comprises multiple components, it is intuitive to allocate weights to each individual element of the PDE loss and dynamically optimize them. Thus, in this study, we propose an extended version of the dynamic weights algorithm where we introduce weights for each component of the PDE loss in the composite PDE loss:
\begin{align}
\label{eq:loss_ex}
    L &= \alpha L_{b} + L_{c} + \gamma L_{x} + \zeta L_{z} \\ 
\label{eq:ex_dw}
    \hat{\alpha}^{k+1} &= \frac{\overline{|{\nabla}_{\theta} L_{c}|}}{\overline{|{\nabla}_{\theta} L_{b}|}}, \:\:\:\;
    \hat{\gamma}^{k+1} = \frac{\overline{|{\nabla}_{\theta} L_{c}|}}{\overline{|{\nabla}_{\theta} L_{x}|}}, \:\:\:\;
    \hat{\zeta}^{k+1} = \frac{\overline{|{\nabla}_{\theta} L_{c}|}}{\overline{|{\nabla}_{\theta} L_{z}|}} \\
 \label{eq:ex_alpha}
    \alpha^{k+1} &= (1-\lambda)\alpha^{k} + \lambda\hat{\alpha}^{k+1}\\
\label{eq:ex_gamma}
    \gamma^{k+1} &= (1-\lambda)\gamma^{k} + \lambda\hat{\gamma}^{k+1}\\
\label{eq:ex_zeta}
    \zeta^{k+1} &= (1-\lambda)\zeta^{k} + \lambda\hat{\zeta}^{k+1}\\
\end{align}
The loss weights are initialized at $\alpha = 100$, $\gamma = 1$, and $\zeta = 1$. Furthermore, $\lambda$ is set to 0.1 for all the cases. A $\mathtt{tanh}$ nonlinear activation is applied at each hidden layer. The entire PINN code is implemented using PyTorch libraries. For training the PINN, the ADAM optimizer with a learning rate of $5 \times 10^{-4}$ is utilized for 10000 epochs. Additionally, the Xavier initialization technique is employed to initialize neural network parameters at the start of the training process.  

To compare PINN predictions with 2D RANS simulations, two different error metrics are employed: Root Mean Square Error (RMSE) and $l_{2}$ relative error ($\epsilon_{v}$), which represents a Euclidean norm error. The definitions of these metrics are provided in Equations~\ref{eq:rmse} and \ref{eq:l2 relative}.
\begin{equation}
     \label{eq:rmse}
     RMSE = \frac{1}{N} {\parallel V - \hat{V} \parallel}_{2}
\end{equation}
\begin{equation}
     \label{eq:l2 relative}
     \epsilon_{v} \; (\%)= \frac{{\parallel V - \hat{V} \parallel}_{2}}{{\parallel V \parallel}_2} \times 100
\end{equation}
where $V$ is the target or ground truth value of the variable and $\hat{V}$ is the prediction of the variable using PINN.

\section{Results}

\subsection{Effect of Non-Dimensionalization}
\begin{figure}[htbp]
    \centering
    \includegraphics[clip, trim=3.5cm 4cm 4cm 4cm,scale=0.26]{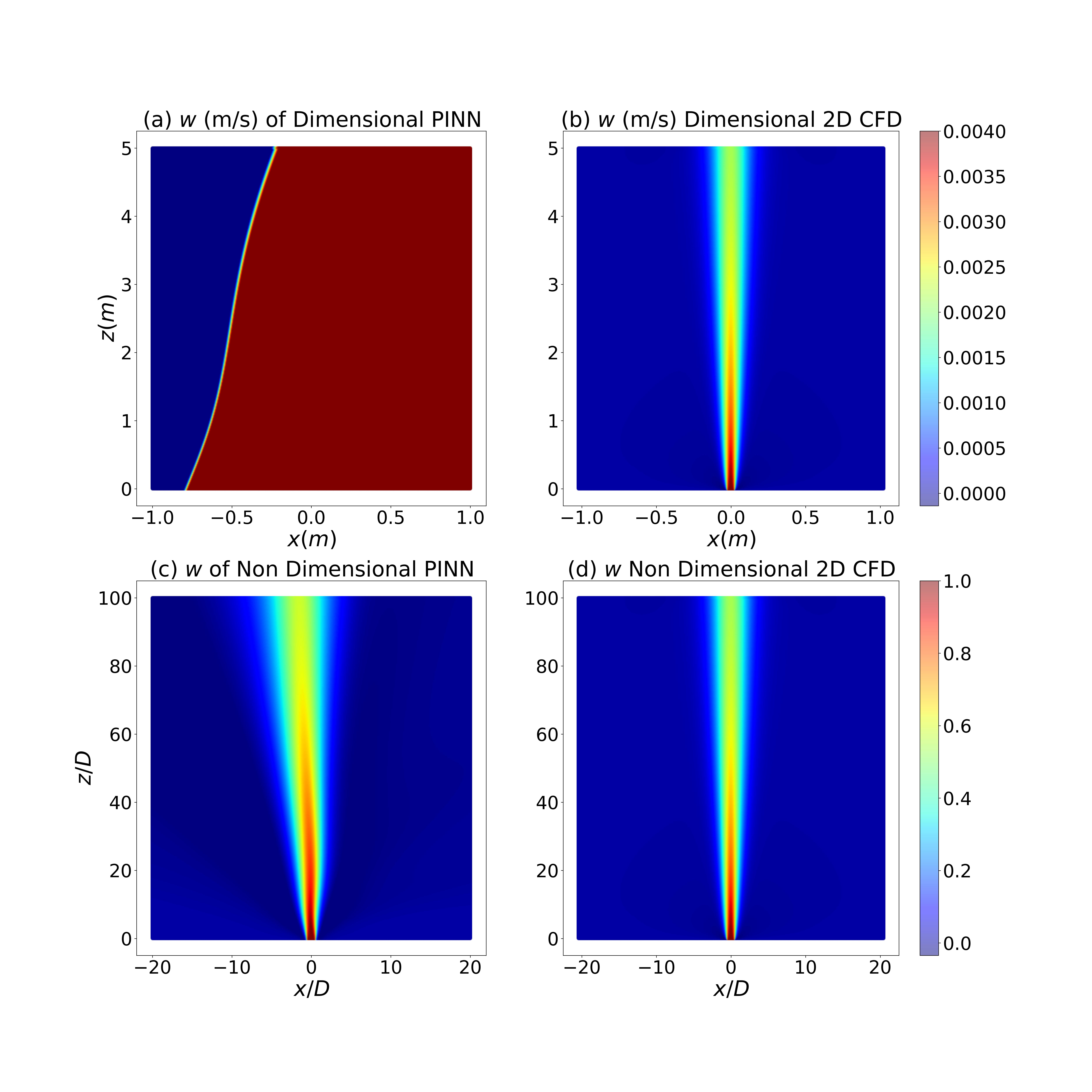}
    \caption{Plots of (a) Dimensional PINN $w$ prediction (b) Dimensional 2D CFD $w$ prediction, and (c) Non-Dimensional PINN $w$ prediction (d) Non-Dimensional 2D CFD $w$ prediction for laminar jet flow at $Re = 200$.} 
    \label{fig:non-dim-w}
\end{figure}

Previous studies on PINNs applied to fluid flow problems used both dimensional and non-dimensional forms of the Navier-Stokes equations. For example, Chengping et.\ al.\ \cite{RAO2020207} used the dimensional Navier-Stokes equations to simulate the flow past the cylinder using PINN with an accuracy of 1.8\% $l_2$ relative error. On the other hand, Jin et.\ al.\ \cite{JIN2021109951} used non-dimensional Navier-Stokes equations to simulate Kovasznay flow, 2D cylinder wake, and 3D Beltrami flow using PINN with an accuracy of 0.08\% $l_2$ relative error. Thus, it is necessary to compare the ability of PINN to approximate the non-dimensional Navier-Stokes and standard Navier-Stokes equations. The dimensional form of Navier-Stokes equations are given in Equations~\ref{eq:NS_C},\ref{eq:NS_X}, and \ref{eq:NS_Z}. The non-dimensional form of the Navier-Stokes equation is given in Equations~\ref{eq:NS_ND_C},\ref{eq:NS_ND_X}, and \ref{eq:NS_ND_Z}. To compare the effect of non-dimensionalization, we considered the architecture given in Fig~\ref{fig:pinn_laminar} with $100 \times 10$ hidden layers with extended dynamic weights for the laminar jet. The inlet velocity in this case is $w_{i} = 0.004$ m/s which corresponds to a Reynolds number of $Re=200$. The PINN was run considering both dimensional and non-dimensional equations for 10,000 epochs with $\lambda = 0.1$, $\alpha = 100$, $\gamma = 1$, and $\zeta = 1$ as the initial weights. The results of the two simulations are shown in Fig~\ref{fig:non-dim-w}.
\begin{align}
\label{eq:NS_C}
    \frac{\partial u}{\partial x} + \frac{\partial w}{\partial z} \; &= \; 0, \\
\label{eq:NS_X}
    u\frac{\partial u}{\partial x} + w\frac{\partial u}{\partial z} &= -\frac{1}{\rho}\frac{\partial p}{\partial x} + \nu\brc{\frac{\partial^2 u}{\partial x^2} + \frac{\partial^2 u}{\partial z^2}}, \\
\label{eq:NS_Z}
    u\frac{\partial w}{\partial x} + w\frac{\partial w}{\partial z} &= -\frac{1}{\rho}\frac{\partial p}{\partial z} + \nu\brc{\frac{\partial^2 w}{\partial x^2} + \frac{\partial^2 w}{\partial z^2}}.
\end{align}

Observing Fig~\ref{fig:non-dim-w} suggests that the non-dimensional variant of the PINN tends to outperform its dimensional counterpart. Notably, the dimensional form appears to face challenges in capturing the jet phenomenon. It can be inferred that the non-dimensionalization acts analogous to normalization used in machine learning to enhance the training procedure and therefore improves the performance of the PINN.

\subsection{Effect of loss weights}
The significance of effectively tuning the weights for each loss component in training PINNs becomes apparent when dealing with diverse flow fields. Traditional PINNs assume larger weights for boundary and initial condition losses and keep these weights constant during the training process. However, the static weights in PINNs might not yield satisfactory results. To address this issue, the method of dynamically tuning the loss weights was introduced by \cite{wang2021understanding}, which optimizes the loss weights during each epoch using the loss gradients. 

The dynamic weights method was used only to optimize the boundary and initial condition loss weights. However, the PDE loss also has various components such as x-momentum, z-momentum, and continuity losses.  Hence, it is pertinent to introduce dynamic weights to individual components of PDE losses as well. Thus, we formulated a novel extended version of dynamic weights for individual PDEs while keeping the continuity equation loss as a base loss for Navier-Stokes/RANS equations.

\begin{figure}[!h]
    \centering
    \includegraphics[clip, trim=3cm 4cm 4cm 4cm,scale=0.27]{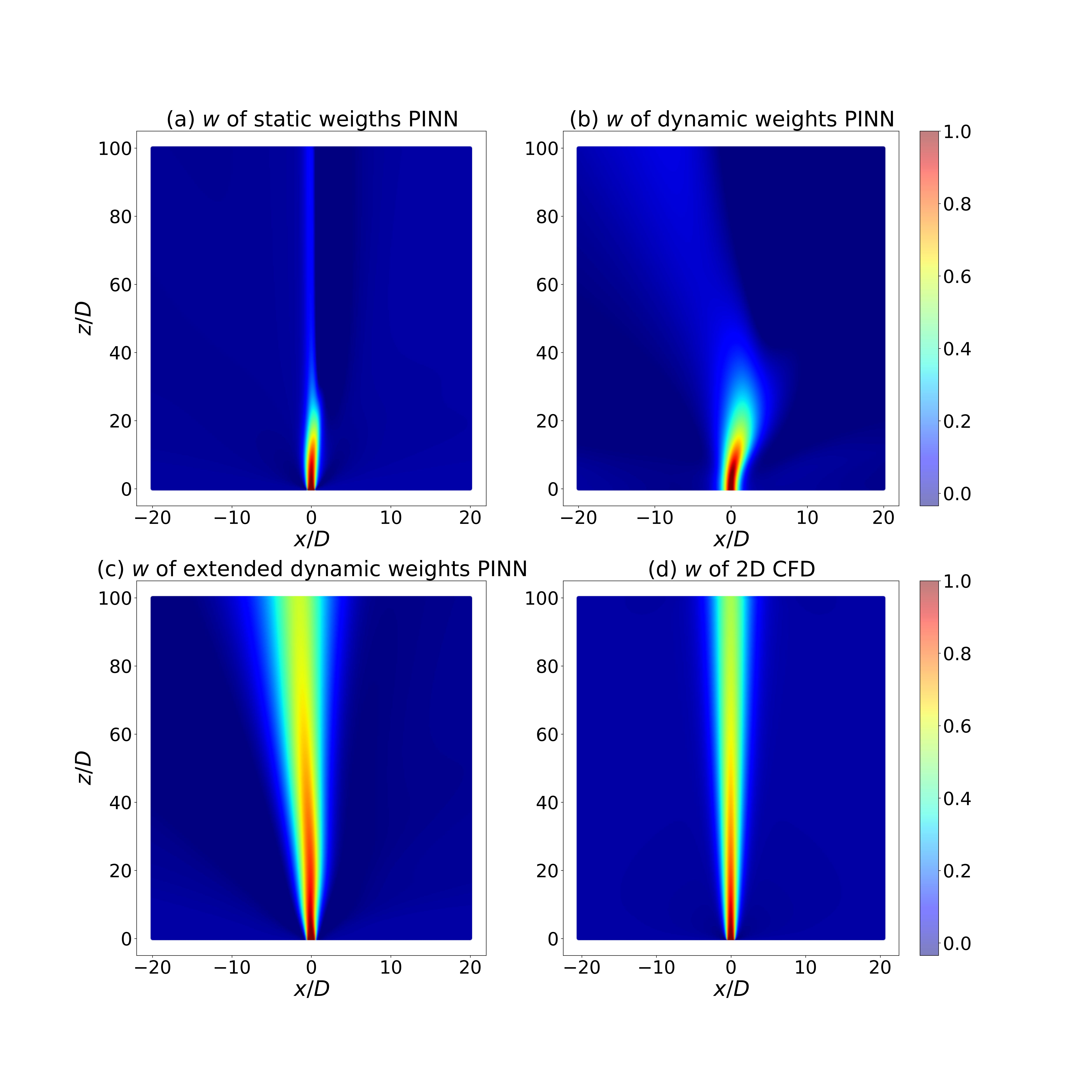}
    \caption{Plots of (a) static-weights-PINN $w$ prediction, (b) dynamic-weights-PINN $w$ prediction, and (c) extended-dynamic-weights-PINN $w$ prediction (d) 2D CFD $w$ prediction for laminar jet flow at $Re = 200$} 
    \label{fig:loss-weights-w}
\end{figure}

To assess the impact of the dynamic weights method, the non-dimensional form of the Navier Stokes equations was used with PINN for a laminar jet at $Re=200$. The simulation was run for 10,000 epochs with the architecture of $100 \times 10$. In the static method, the loss weights were kept the same for all epochs: $\alpha = 100$, $\gamma = 1$, and $\zeta = 1$. For the dynamic and extended dynamic weights, these values of weights were taken as the initial values which were then optimized with the help of loss gradients with $\lambda = 0.1$. The results of different formulations of loss weights for predicting $w$ are illustrated in Fig~\ref{fig:loss-weights-w}. 

 From Fig~\ref{fig:loss-weights-w} and Table~\ref{tab:loss-weights-errors}, it can be seen that the extended dynamic weights are performing better than the static and dynamic weights counterparts. This can be attributed to the fact that it is necessary to distribute the loss gradients of individual PDEs as well, apart from boundary and initial condition losses.

  \begin{figure}[!h]
    \centering
    \includegraphics[clip, trim=2cm 0cm 2cm 0cm,scale=0.27]{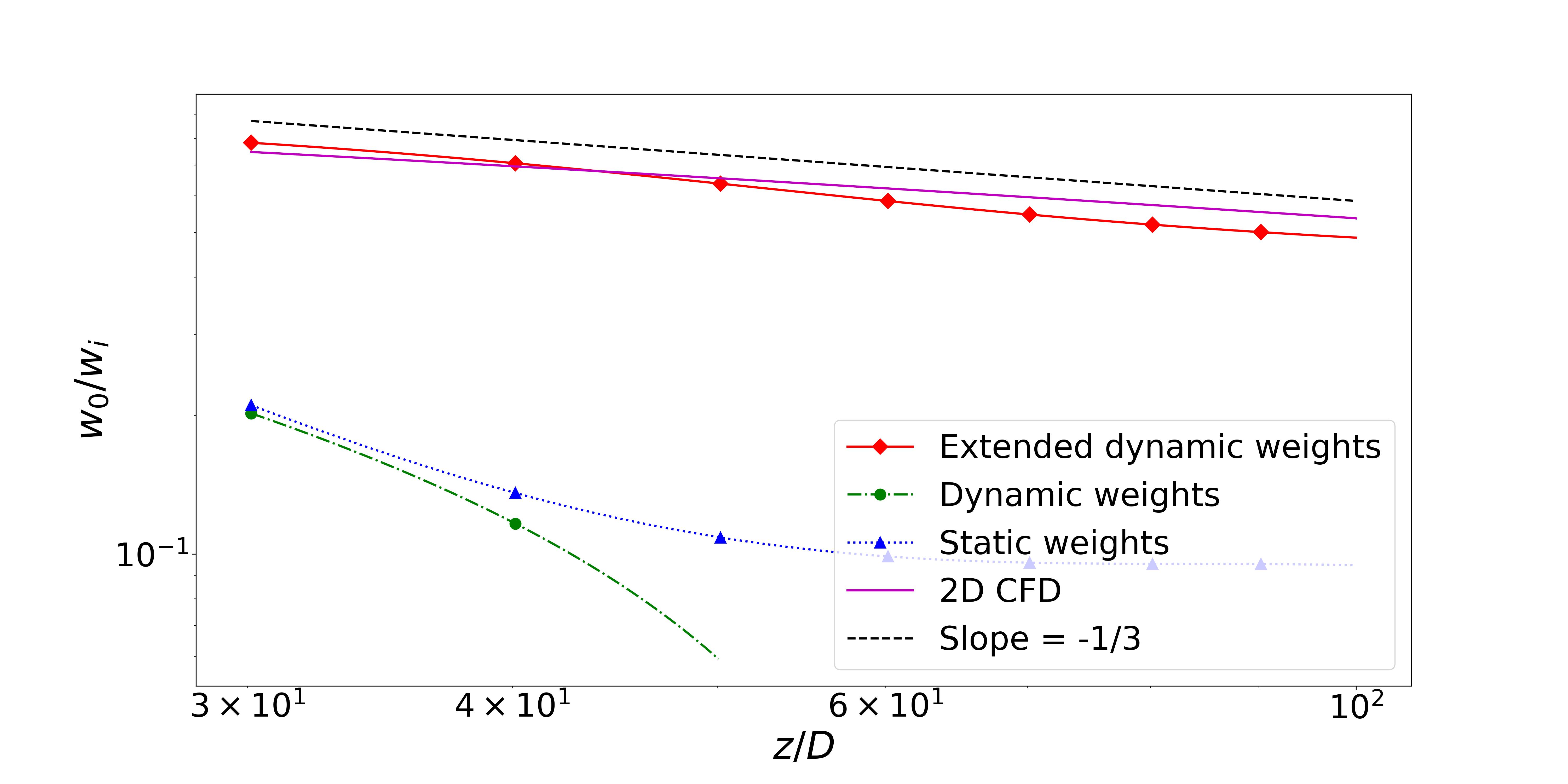}
    \caption{Comparison between the center-line $w$ velocity decay for different loss weights formulation with the 2D CFD prediction for laminar jet flow at $Re = 200$ beyond $z = 30 \times D$} 
    \label{fig:loss_weights_cl}
\end{figure}

 \begin{figure}[!h]
    \centering
    \includegraphics[clip, trim=2cm 0cm 2cm 0cm,scale=0.27]{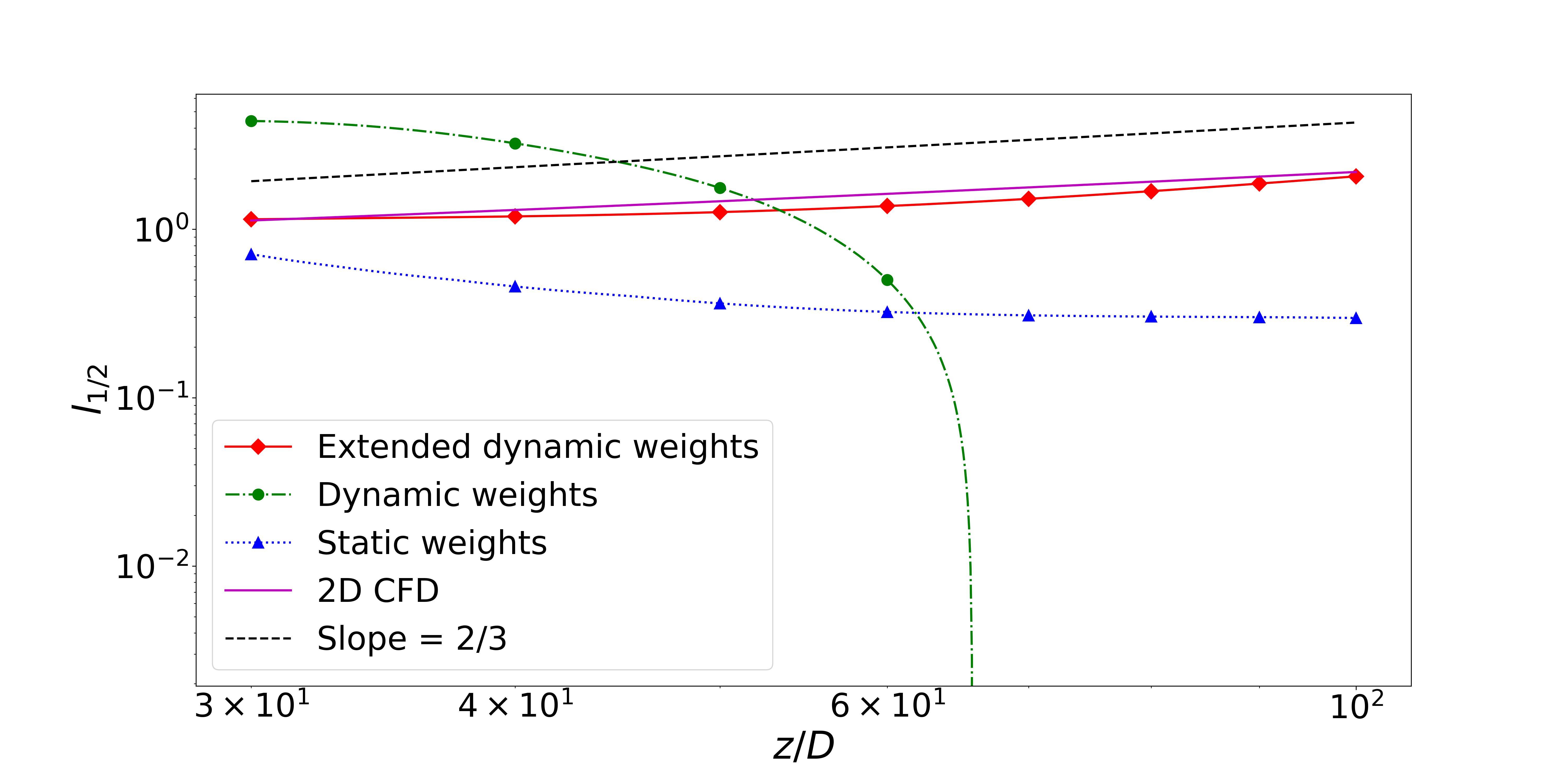}
    \caption{Comparison between the jet half-width for different loss weights formulation with the 2D CFD prediction for laminar jet flow at $Re = 200$ beyond $z = 30 \times D$} 
    \label{fig:loss_weights_js}
\end{figure}

To further support the claim that the extended dynamic weights perform better than their counterparts, the center-line velocity decay is shown in Fig~\ref{fig:loss_weights_cl}, and the half-width of the jet is shown in Fig~\ref{fig:loss_weights_js}. It can be seen that the PINN results using extended dynamic weights closely agree with the CFD results.

 \begin{table}[!h]
  \caption{Comparison of errors of different flow variables between different loss weights formulations for laminar jet flow at $Re = 200$ beyond $z = 30 \times D$}
  \begin{center}
  \begin{tabular}{ | c | c | c | c | c | c | c | }
    \hline
    \multirow{2}{4em}{\textbf{Variables}}  &  \multicolumn{2}{c |}{\textbf{Static Weights}} & \multicolumn{2}{c |}{\textbf{Dynamic Weights}} & \multicolumn{2}{c |}{\textbf{Extended Dynamic Weights}}\\
    \cline{2-7}
    & RMSE & $\epsilon_v (\%)$ & RMSE &$\epsilon_v (\%)$ & RMSE & $\epsilon_v (\%)$ \\
    \hline 
    $w_0$ & 0.51 & 82 & 0.599 & 96.3 & 0.041 & 6.6 \\
    $l_{1/2}$ & 1.381 & 80.6 & 1.309 & 224.3 & 0.2 & 11.6 \\
    \hline
  \end{tabular}
  \label{tab:loss-weights-errors}
  \end{center}
\end{table}
\subsection{Effect of Architectures on Turbulent Jet}
So far we have discussed the laminar 2D planar jet and the challenges in predicting flow variables using PINN. In the subsequent discussions, we consider turbulent 2D planar jets. Reynolds-Averaged Navier Stokes (RANS) equations are considered in this study to predict mean flow variables using PINN architecture. However, the RANS equations have their own set of challenges such as the Reynolds stress closure problem. The Reynolds stresses in RANS equations are unknown and require additional modeling. Several turbulence models are currently in use to model the Reynolds stresses, and the degree of complexity across different turbulence models varies depending on the problem being considered.  \\

In this study, to model the Reynolds stresses, different PINN configurations were considered as described above. Four different architectures, A1, A2, A3 and A4, are considered and results from these configurations are compared with the CFD results to assess their accuracy. 


From the results of the laminar jet, it was evident that the extended dynamic weights formulation for loss weights was more accurate when compared to static and dynamic loss weights formulation. Hence, extended dynamic weights formulation for loss weights is used for all cases of turbulent jet. To simulate the turbulent jet, PINN was run for 10,000 epochs with $\lambda = 0.1$ and loss weights of $\alpha = 100$, $\gamma = 1$ and $\zeta = 1$ as initial loss weights for A1 and A2 configurations.  For A4 architecture, PINN was run for 7000 epochs with $\lambda = 0.01$ due to high oscillations observed during convergence towards the end.

\begin{figure}[!h]
    \centering
    \includegraphics[clip, trim=8cm 0.5cm 4cm 1.5cm,scale=0.2]{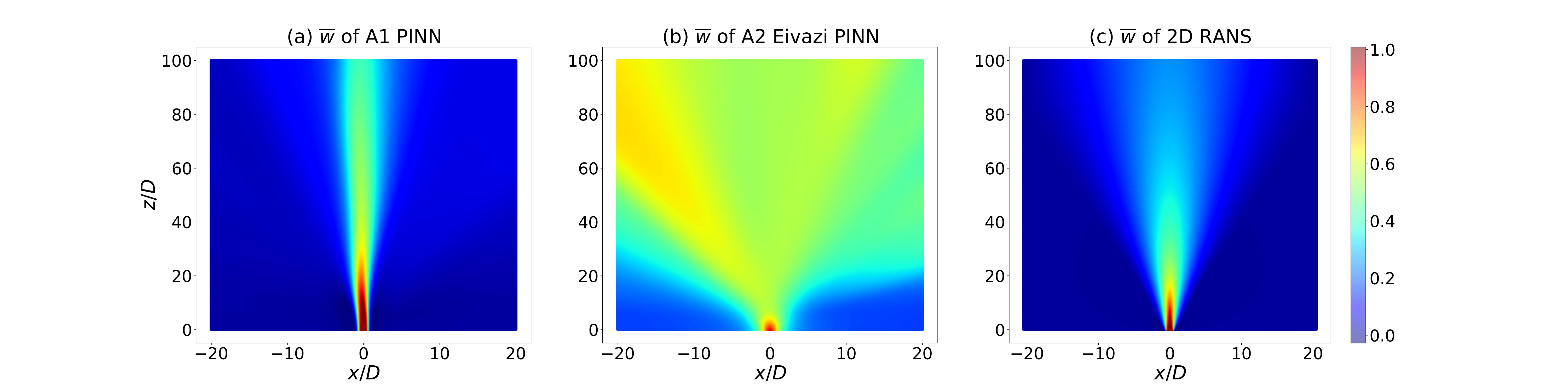}
    \caption{Comparison between the (a) A1 PINN $\overline{w}$ prediction, and (b) A2 PINN $\overline{w}$ prediction from Eivazi \cite{eivazi2022physics} with the (c) 2D CFD $\overline{w}$ prediction for turbulent jet flow at $Re = 5000$} 
    \label{fig:turb_ml-w}
\end{figure}

In the first architecture A1, a simple one-equation turbulence model is considered, namely the mixing length model, to predict the dominant component of the Reynolds stress $\overline{u'w'}$ as a function of mixing length and velocity gradients. Along with mean velocity components and mean pressure, mixing length is also one of the outputs of PINN as shown in Fig~\ref{fig:pinn_turbulent_ml}. The mean axial velocity predicted by A1 is shown in Fig~\ref{fig:turb_ml-w} (a). Further, the respective errors of different flow variables are given in Table~\ref{tab:turb-ml}.

\begin{table}[htbp]
  \caption{Comparison of errors of different flow variables between A1, and A2 of PINN for the turbulent jet at $Re = 5000$ beyond $z = 30 \times D$}
  \begin{center}
  \begin{tabular}{ | c | c | c | c | c | }
    \hline
    \multirow{2}{4em}{\textbf{Variables}} &  \multicolumn{2}{c |}{\textbf{A1}} & \multicolumn{2}{c |}{\textbf{A2}} \\
    \cline{2-5}
    & RMSE & $\epsilon_v (\%)$ & RMSE & $\epsilon_v (\%)$ \\
    \hline 
    $\overline{w}_0$ & 0.175 & 52.7 & 0.226 & 68.3 \\
    $\overline{l}_{1/2}$ & 4.198 & 58.6 & 7.163 & 100 \\
    \hline
  \end{tabular}
  \label{tab:turb-ml}
  \end{center}
\end{table}

 Different turbulence models have their shortcomings, such as the inability of the $k-\epsilon$ model to predict the slip wall conditions, and may provide different levels of accuracy in results depending on the type of flow considered. To circumvent this problem in turbulence modeling, Eivazi et.\ al.\ \cite{eivazi2022physics}  proposed to predict the Reynolds stresses directly from PINN as shown in Fig~\ref{fig:pinn_RANS}. We considered this method in A2, where all the 3 components of Reynolds stresses are outputs of the PINN. The mean axial velocity obtained from this PINN configuration is shown in Fig~\ref{fig:turb-w} (b). The predictions of mean flow variables seem to be poor when compared to results from A1. 
This phenomenon may be attributed to the non-closure issue, characterized by three equations and six unknowns that need to be resolved simultaneously.

\begin{figure}[!h]
    \centering
    \includegraphics[clip, trim=8cm 0.5cm 4cm 1.5cm,scale=0.2]{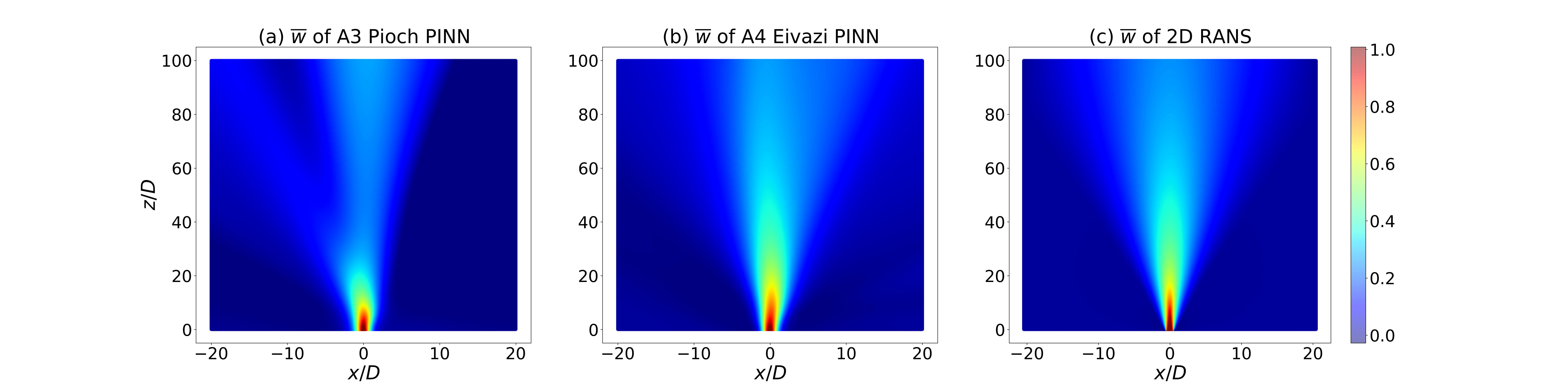}
    \caption{Comparison between the (a) A3 PINN $\overline{w}$ prediction from Pioch \cite{pioch2023turbulence}, and (b) Proposed A4 PINN $\overline{w}$ prediction with the (c) 2D CFD $\overline{w}$ prediction for turbulent jet flow at $Re = 5000$} 
    \label{fig:turb_nu_t-w}
\end{figure}

 To further improve the results from A2, Pioch et.\ al.\ \cite{pioch2023turbulence} proposed to leverage the Boussinesq hypothesis, given Equation~\ref{eq:bh}, and predict the turbulent viscosity $\nu_t$ as a part of PINN outputs in the A3 configuration, as shown in Fig~\ref{fig:pinn_nut}. The mean axial velocity obtained from this PINN configuration is shown in Fig~\ref{fig:turb_nu_t-w} (a). The predictions of mean flow variables seem to be better when compared to results from A1 and A2. Nevertheless, despite these modifications, the non-closure issue persists in the A3 configuration, indicating the necessity for further refinements to achieve accurate results.

\begin{table}[htbp]
  \caption{Comparison of errors of different flow variables between A3, and A4 of PINN for the turbulent jet at $Re = 5000$ beyond $z = 30 \times D$}
  \begin{center}
  \begin{tabular}{ | c | c | c | c | c | }
    \hline
    \multirow{2}{4em}{\textbf{Variables}} &  \multicolumn{2}{c |}{\textbf{A3}} & \multicolumn{2}{c |}{\textbf{A4}} \\
    \cline{2-5}
    & RMSE & $\epsilon_v (\%)$ & RMSE & $\epsilon_v (\%)$ \\
    \hline 
    $\overline{w}_0$ & 0.102 & 31.1 & 0.008 & 2.3 \\
    $\overline{l}_{1/2}$ & 1.951 & 27.2 & 1.921 & 26.8 \\
    \hline
  \end{tabular}
  \label{tab:turb-nut}
  \end{center}
\end{table}

Finally, in the fourth configuration A4, we propose a novel technique where two separate neural networks are considered. In the first neural network, mean velocities and pressure are the outputs. Further, the auto-differentiation technique is used to compute the velocity gradients which act as inputs to the second neural network. Turbulent viscosity $\nu_t$ is the output of the second neural network and the Boussinesq hypothesis is used to finally compute the Reynolds stresses. The novel PINN architecture proposed in A4 is shown in Fig~\ref{fig:pinn_2NN_nut}. In A4, the second neural network serves as a means to address the turbulent viscosity closure, thereby enhancing the predictive capabilities of the PINN. The mean axial velocity obtained from this configuration is shown in Fig~\ref{fig:turb-w} (b) and the respective errors are given in Table~\ref{tab:turb-nut}.

\begin{figure}[!h]
    \centering
    \includegraphics[clip, trim=0cm 0cm 2cm 0cm,scale=0.23]{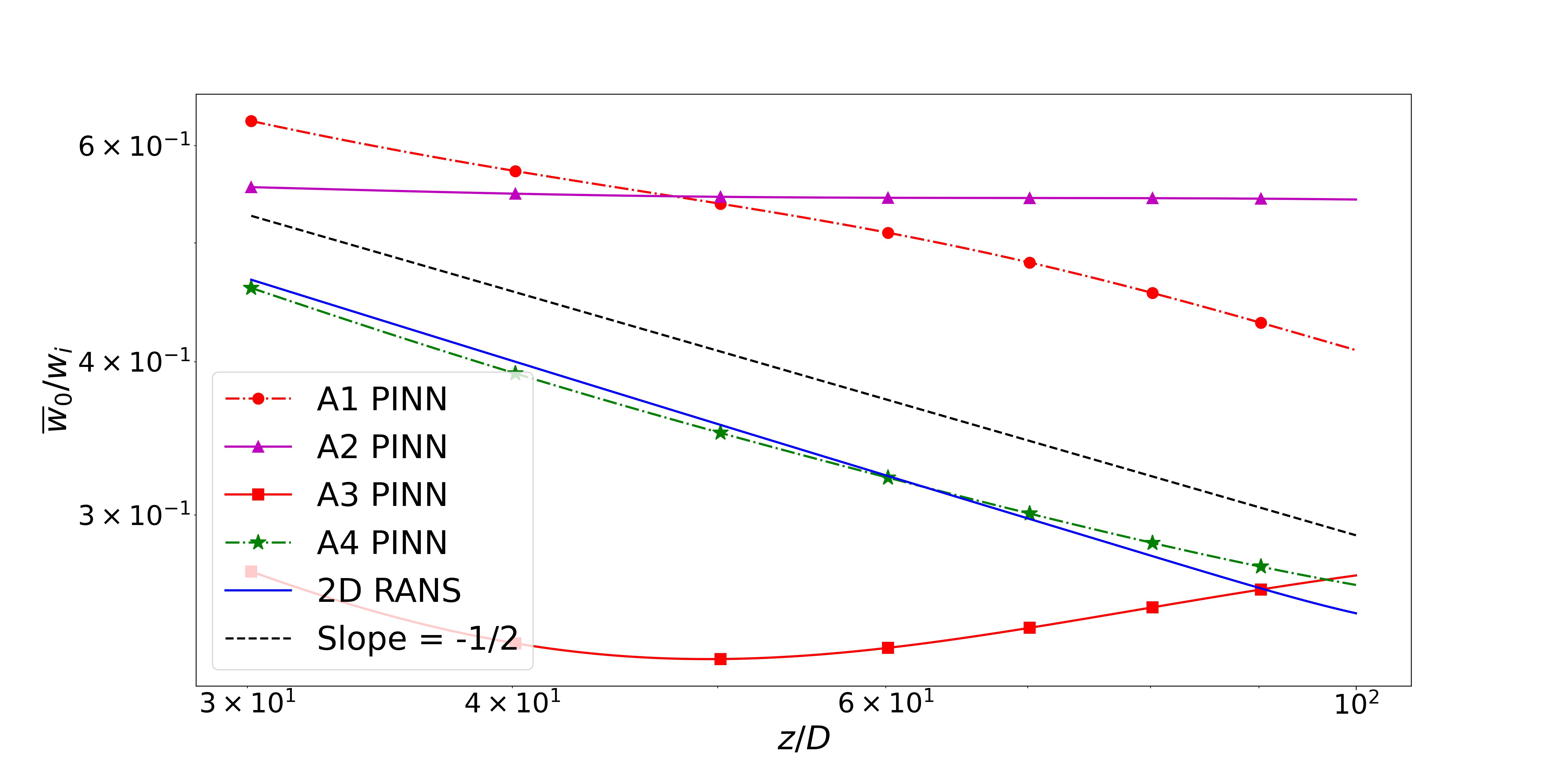}
    \caption{Comparison between the center-line axial velocity for different cases of PINN with the 2D RANS prediction for turbulent jet flow at $Re = 5000$ beyond $z = 30 \times D$} 
    \label{fig:turb_cl}
\end{figure}

\begin{figure}[!h]
    \centering
    \includegraphics[clip, trim=2cm 0cm 2cm 0cm,scale=0.23]{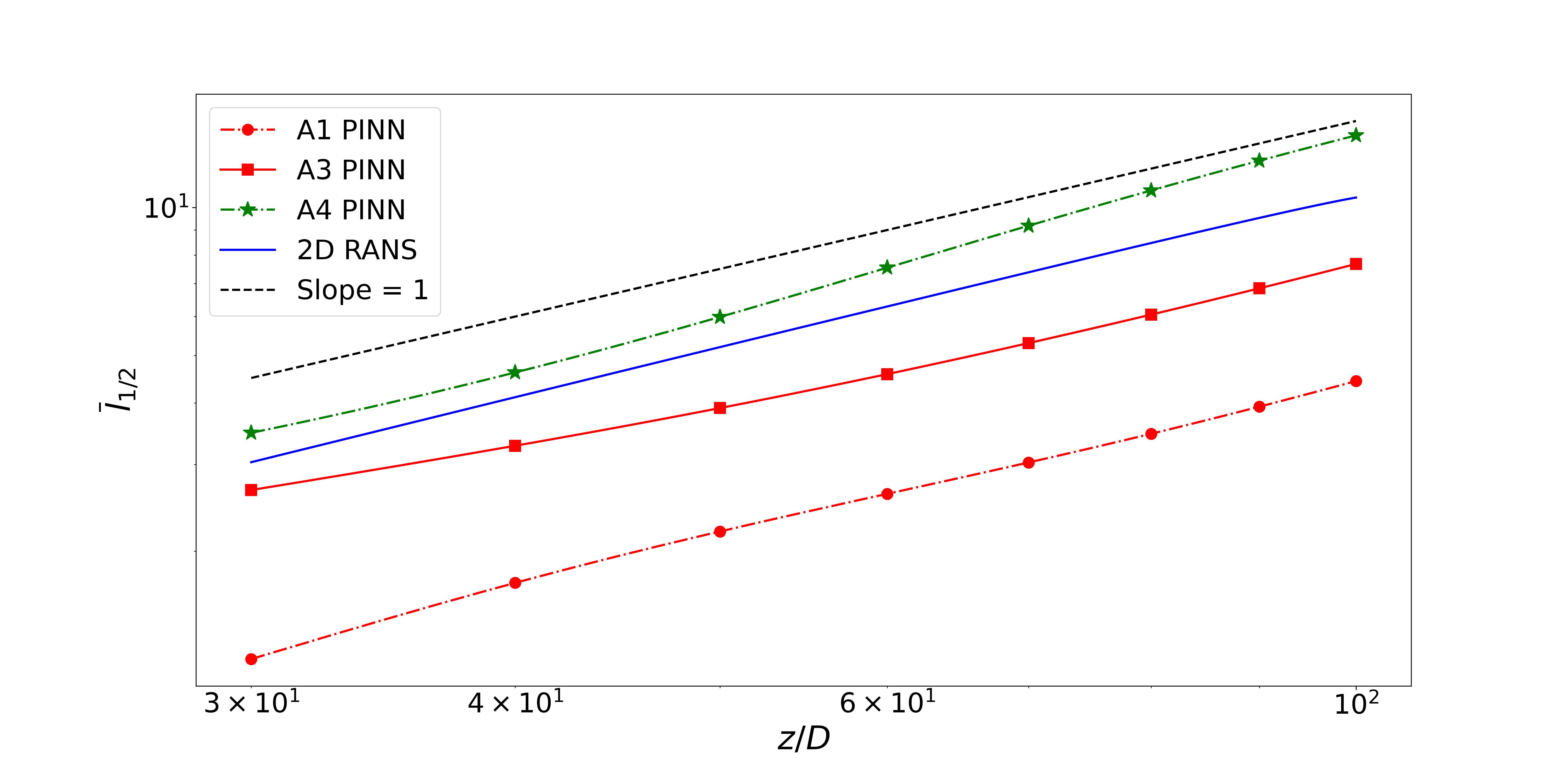}
    \caption{Comparison between the jet half-width for different cases of PINN with the 2D RANS prediction for turbulent jet flow at $Re = 5000$ beyond $z = 30 \times D$} 
    \label{fig:turb_hw}
\end{figure}

 From Fig~\ref{fig:turb_ml-w}, and \ref{fig:turb_nu_t-w} and Table~\ref{tab:turb-ml}, and \ref{tab:turb-nut}, it can be seen that the proposed A3 architecture works better compared to configurations A1 and A2. Qualitatively, the contours of mean axial velocity are very similar to the CFD result. Also, centerline velocity decay and half-width of the jet match closely with the CFD result in the self-similar region $z/D>30$ as shown in figures \ref{fig:turb_cl} and \ref{fig:turb_hw}, respectively. From the error metrics shown in Table~\ref{tab:turb-ml}, and \ref{tab:turb-nut}, it is evident that A4 has the least error and closest agreement with the CFD result.

\pagebreak

\section{Discussion and Summary}
This work is focused on predicting the mean statistics of a turbulent jet using the PINN with various configurations and neural network architectures. In the first part, the laminar jet problem was considered using various configurations. To begin with, a comparison was made between the predictions obtained from PINN using dimensional and non-dimensionalized Navier-Stokes equations as loss functions. We found that PINN with non-dimensional equations significantly improved the flow prediction. The effect of loss weights is also considered by dynamically varying the weights of boundary as well as PDE losses. Further, we introduced a new technique called extended dynamic weights where weights for each component of the PDE loss are also varied dynamically. We found a significant improvement in the results using the proposed extended dynamics weights technique. In the second part of the study, the turbulent jet was considered using the non-dimensional equations along with the extended dynamic weights technique. Further, we proposed a novel two-neural network architecture for predicting the turbulent viscosity to address the closure problem without assuming any underlying turbulence models.

The major conclusions drawn from this study are as follows:
\begin{itemize}
    \item Non-dimensionalizing the inputs, outputs, and PDEs greatly helps the PINN to accurately predict the flow field since it acts as normalization in machine learning that is used to boost the training performance of the model.

    \item The weights of different losses play a significant role in predicting the flow. In this study, to overcome the problem of gradient pathology faced by PDEs, we proposed an extended version of dynamic weights for the PDEs which improved the prediction of mean flow statistics significantly. 
    
    \item For turbulent jets, we proposed a novel architecture where two neural networks are used. The first one predicts mean flow velocity components and mean pressure which act as inputs to the second neural network. Turbulent viscosity $\nu_t$ is the output of the second neural network and finally, the Boussinesq hypothesis is used to compute the Reynolds stress components. 
    
    \end{itemize}

With the proposed novel techniques of extended dynamic weights and two neural networks in conjunction with the Boussinesq hypothesis, we were able to predict the flow statistics of laminar and turbulent jets accurately. However, there is scope for further improvement. The errors in the proposed architecture A4 primarily stem from the implementation of boundary conditions and pressure field predictions which cause the jet to be asymmetric about the center-line. Future studies can focus on the implementation of pressure-velocity coupling in the PINN architecture to further improve the predictions. Also, the generalizability of the proposed PINN architecture is still to be explored, to whether the proposed architecture will work for different sets of boundary conditions and other fluid flows will be investigated in future studies.

\bibliographystyle{unsrtnat}

\bibliography{main}

\end{document}